# Accuracy of the Muskingum-Cunge method for constant-parameter diffusion-wave channel routing with lateral inflow

Li Wang [a], Sergey Lapin [c,d,*], Joan Q. Wu [b], William J. Elliot [e], Fritz R. Fiedler [f]

[a] Yakama Nation ERWM, Union Gap, WA 98903, USA

[b] Washington State University, Department of Biological Systems Engineering, Puyallup Research and Extension Center, Puyallup, WA 98371, USA

[c] Washington State University, Department of Mathematics and Statistics, Pullman, WA 99164, USA

[d] Kazan Federal University, Institute of Computational Mathematics and IT, Kazan, Russia

[e] US Department of Agriculture, Forest Service, Rocky Mountain Research Station, Moscow, ID 83843, USA

[f] University of Idaho, Department of Civil Engineering, Moscow, ID 83844, USA

[*] Corresponding author

E-mail address: slapin@wsu.edu (S. Lapin).

**Abstract**

Channel routing is important in flood forecasting and watershed modeling. The general constant-parameter Muskingum-Cunge (CPMC) method is second-order accurate and easy to implement. With specific discretizations such that the temporal and spatial intervals maintain a unique relationship, the CPMC method can be third-order accurate. In this paper, we derive the average lateral inflow term in the second- and third-order accuracy CPMC method, and demonstrate that For spatially and temporally variable lateral inflow, the effect of lateral inflow on simulated discharge varies with spatial and temporal discretizations, the value and spatial and temporal variations of lateral inflow, wave celerity, and diffusion coefficient. Comparison of the CPMC solution with the analytical solution shows that both the second- and third-order accuracy schemes are more accurate than the simplified method by which spatial derivatives of lateral inflow are ignored. For small time steps, the third-order accuracy CPMC method results in higher accuracy than the second-order scheme even when the third-order accuracy criterion is not fully met. For large time steps, the temporal and spatial discretization of the third- and second-order scheme becomes the same, but the third-order scheme yields higher accuracy than the second-order scheme because of the third-order accurate estimation of the lateral inflow term.

## 1. Introduction

Channel upstream inflow is usually the most important component for flood routing. In watershed modeling, however, channel water often comes from lateral inflow. As in the Water Erosion Prediction Project (WEPP) model, water generated from a hillslope (surface runoff, subsurface lateral flow, and groundwater base flow) may enter a channel as upstream inflow when the hillslope is at the top of the channel, or as lateral inflow when the hillslope is on the side of the channel (Fig. 1).

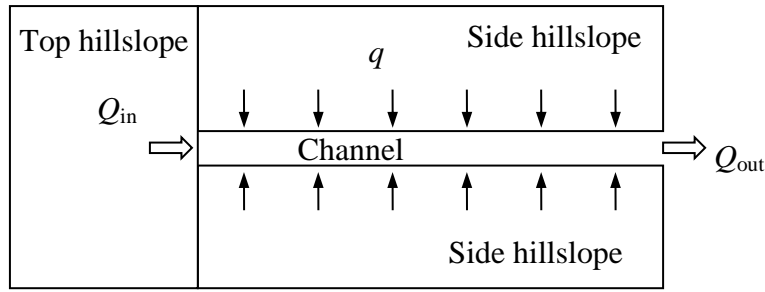

Figure 1. Schematic of the relationship between hillslopes and a channel segment in the Water Erosion Prediction Project (WEPP) model.

A hillslope can be at the top of a channel only in cases of 1st-order channels, and would otherwise be on the side of the channel, with the top of the channel being upstream channels or an impoundment (Flanagan and Livingston, 1995). In addition, the gain or loss of the stream water by precipitation, infiltration, and evapotranspiration is often included in the lateral inflow term. In commonly used numerical channel routing methods, e.g., the Muskingum-Cunge or the kinematic-wave method, we need to calculate the average lateral inflow in the channel routing equation. The order of accuracy of the average lateral inflow term can be a dominant factor affecting the accuracy of the numerical channel routing in watershed simulations.

Price (2009) developed a second-order accurate nonlinear diffusion-wave scheme and solved it using the Newton-Raphson iterative method. The author also analyzed the effect of bed slope on the accuracy and found the accuracy decreased with decreasing bed slope. However lateral inflow was considered to be uniformly distributed in his study. Todini (2007) studied variable parameter Muskingum-Cunge (VPMC) method and developed a mass-conservative approach by resolving the storage and steady-state inconsistencies of the original VPMC method. In his study, lateral inflow was not considered. Barry and Bajracharya (1995) showed that for channel routing without lateral inflows, when the time step $\Delta t$ and the space interval $\Delta x$ maintained a certain relationship so that the Courant number is 0.5, the Muskingum-Cunge method was third-order accurate. For constant-parameter diffusion-wave channel flows without lateral inflow, Bajracharya and Barry (1997) derived a relationship of spatial and temporal steps of $\Delta t = \dfrac{C\Delta x - 2D}{C^2}$, where $C$ denotes kinematic celerity and $D$ diffusion coefficient, to assure a second-order accuracy of Muskingum-Cunge scheme, a relationship of $\Delta t = \dfrac{1}{C}\sqrt{\Delta x^2 - \dfrac{12D^2}{C^2}}$ or $\Delta x = \sqrt{C^2 \Delta t^2 + \dfrac{12D^2}{C^2}}$ for third-order accuracy, and fixed $\Delta t = \dfrac{2\sqrt{3}D}{C^2}$ and $\Delta x = \dfrac{2\sqrt{6}D}{C}$ for fourth-order accuracy. Szel and Gaspar (2000), without considering lateral inflow, related the temporal and spatial intervals to the Courant number $C_r = \dfrac{C\Delta t}{\Delta x}$ and Peclet number $P_e = \dfrac{C\Delta x}{2D}$, discussed their effect on the stabilities of the Muskingum-Cunge scheme, and found that the relationship of the spatial and temporal steps required for the third-order Muskingum-Cunge method can be simplified to a dimensionless equation $C_r^2 + \dfrac{3}{P_e^2} - 1 = 0$.

In addition to the relationship between $\Delta x$ and $\Delta t$, Moramarco et al. (1999) reported that the choice of reference discharge, which is used to calculate the kinematic celerity and the diffusion coefficient, can also affect the accuracy of the channel routing with lateral inflow. By testing the channel routing without the upstream inflow, they found that the error in channel routing changed with reference discharge and bed slope. For a channel with a relatively gentle slope, such as 0.0001, the selected reference discharge should be larger for a better accuracy; for a channel with a rather steep slope, e.g., 0.01, the accuracy of the channel routing was not sensitive to the reference discharge.

The lateral inflow in a channel routing equation was usually treated as concentrated or uniformly distributed for simplicity (Chow et al., 1988; Fan and Li, 2006). When lateral inflow is spatially and temporally variable, its effect on accuracy of numerical channel routing has not been discussed. In this study, we will (i) derive a second- and third-order accurate representation for the lateral inflow term used in the constant-parameter Muskingum-Cunge (CPMC) method for channel routing, (ii) compare the results from the third- and the second-order accuracy CPMC methods with analytical solution, and analyze the effect of the time-step size on the accuracy of the CPMC solution.

## 2. Methods

The constant-parameter diffusion-wave equation with lateral inflow can be simplified as (Lighthill and Whitham, 1955; Bajracharya and Barry, 1997; Fan and Li, 2006; Price, 2009)

$$\frac{\partial Q}{\partial t} + C\frac{\partial Q}{\partial x} - D\frac{\partial^2 Q}{\partial x^2} = Cq \tag{1}$$

where $Q = Q(x,t)$ is discharge (m$^3$ s$^{-1}$), $x$ is downstream distance (m), $t$ is time (s), $q = q(x,t)$ is lateral inflow rate per unit length (m$^2$ s$^{-1}$), with positive $q$ representing flow into, and negative $q$ flow out of, the channel, $C = \frac{dQ_R}{dA}$ is kinematic wave celerity (m s$^{-1}$), and $D = \frac{Q_R}{2BS_f} \approx \frac{Q_R}{2BS_0}$ is the diffusion coefficient (m$^2$ s$^{-1}$) where $Q_R$ is the reference discharge, $A$ the cross-sectional area (m$^2$), $B$ the channel width at the water surface (m), $S_f$ the friction slope, and $S_0$ the channel bed slope.

### 2.1 Third-order accuracy CPMC method

The Muskingum-Cunge method solving Eq. (1) numerically is (Chow et al., 1988; Ponce, 1995; Bajracharya and Barry, 1997; Szel and Gaspar, 2000)

$$Q_{i+1}^{j+1} = C_1 Q_i^j + C_2 Q_i^{j+1} + C_3 Q_{i+1}^j + C_4 \bar{q}_{i+1}^{j+1} \Delta x \tag{2}$$

where the Muskingum-Cunge coefficients are given by

$$C_1 = \frac{\Delta x + C\Delta t - \dfrac{2D}{C}}{\Delta x + C\Delta t + \dfrac{2D}{C}} \tag{3}$$

$$C_2 = \frac{-\Delta x + C\Delta t + \frac{2D}{C}}{\Delta x + C\Delta t + \frac{2D}{C}} \quad (4)$$

$$C_3 = \frac{\Delta x - C\Delta t + \frac{2D}{C}}{\Delta x + C\Delta t + \frac{2D}{C}} \quad (5)$$

and

$$C_4 = \frac{2C\Delta t}{\Delta x + C\Delta t + \frac{2D}{C}} = 1 - C_3 \quad (6)$$

and $\bar{q}_{i+1}^{j+1}$ is the average lateral inflow. For uniformly distributed lateral inflow, it was calculated as $\bar{q}_{i+1}^{j+1} = \frac{q_{i+1}^{j} + q_{i+1}^{j+1}}{2}$ (Chow et al., 1988; Appendix A).

The CPMC method is second-order accurate without restrictions on temporal and spatial discretizations (Appendix A). To achieve the third-order accuracy without changing the representations of the Muskingum-Cunge coefficients, the spatial and temporal intervals must satisfy the following relationships (Bajracharya and Barry, 1997; Szel and Gaspar, 2000; Appendix A)

$$\Delta x = \sqrt{C^2 \Delta t^2 + \frac{12D^2}{C^2}} \quad (7)$$

for a given $\Delta t$, or

$$\Delta t = \frac{1}{C}\sqrt{\Delta x^2 - \frac{12D^2}{C^2}} \quad (8)$$

with $\Delta x$ fixed. Eqs. (7) and (8) are equivalent to the following dimensionless equation (Szel and Gaspar, 2000)

$$C_r^2 + \frac{3}{P_e^2} - 1 = 0 \quad (9)$$

Hence, for a diffusion wave with $P_e > \sqrt{3}$, the simulated outflow is of a higher-order accuracy if

$$C_r = \sqrt{1 - \frac{3}{P_e^2}} \tag{10}$$

The third-order accuracy average lateral inflow can then be calculated as (Appendix A)

$$\bar{q}_{i+1}^{j+1} = q^j + \frac{1}{2} q_t \Delta t + \frac{1}{2} q_x \left( \Delta x + \frac{2D}{C} \right) + \frac{1}{6} q_{tt} \Delta t^2 + \frac{1}{4} q_{xx} \left( \Delta x^2 - \frac{1}{3} C^2 \Delta t^2 + \frac{2D \Delta x}{C} \right)$$
$$+ \frac{1}{4} q_{xt} \left( \Delta x + \frac{1}{3} C \Delta t + \frac{2D}{C} \right) \Delta t + \frac{1}{4} q_{3x} \left( \Delta x - \frac{1}{3} C \Delta t + \frac{2D}{C} \right) D \Delta t + \frac{1}{6} q_{xxt} D \Delta t^2 + \frac{1}{6} q_{4x} D^2 \Delta t^2$$

(11)

We can also show that, for the second-order accuracy CPMC (Appendix A)

$$\bar{q}_{i+1}^{j+1} = q^j + \frac{1}{2} q_t \Delta t + \frac{1}{2} q_x \left( \Delta x + \frac{2D}{C} \right) + \frac{1}{2} q_{xx} D \Delta t \tag{12}$$

Eqs. (11) and (12) show that, $\bar{q}_{i+1}^{j+1}$ depends not only on lateral inflow and its spatial and temporal variation as well as the spatial and temporal discretization, but also on wave celerity and the diffusion coefficient of the channel flow.

If the spatial variation of lateral inflow is negligible, the third- and second-order accuracy average lateral inflow can also be estimated from a discrete dataset (Appendix A), i.e.,

$$\bar{q}_{i+1}^{j+1} = \begin{cases} \dfrac{-q_{i+1}^{j+2} + 8q_{i+1}^{j+1} + 5q_{i+1}^{j}}{12} + O(\Delta t^3) & \text{for } j = 0 \\ \dfrac{5q_{i+1}^{j+1} + 8q_{i+1}^{j} - q_{i+1}^{j-1}}{12} + O(\Delta t^3) & \text{for } j = 1,2,\text{K} \end{cases} \tag{13}$$

for third-order accuracy, and

$$\bar{q}_{i+1}^{j+1} = \frac{q_{i+1}^{j+1} + q_{i+1}^{j}}{2} + O(\Delta t^2) \text{ for } j = 0,1,2\text{K} . \tag{14}$$

for second-order accuracy.

## 2.2 A numerical experiment

To test the accuracy of the CPMC method, we consider a synthetic channel flow

$$Q(x,t) = 2 + \sin\frac{2\pi x}{L} + \sin\frac{2\pi t}{T} \quad (\text{m}^3 \text{ s}^{-1}) \tag{15}$$

for $0 \leq x \leq L$ and $0 \leq t \leq T$

with $L$ = 10,000 m and $T$ = 10,000 s. The width of the rectangular channel is 2 m, bed slope 0.01 so that the effect of the slope steepness on reference discharge can be neglected (Moramarco et al., 1999), and, Manning's roughness coefficient 0.035.

The minimum inflow from Eq. (15), $Q_b$ = 1 m$^3$ s$^{-1}$, and the peak inflow $Q_p$ = 3 m$^3$ s$^{-1}$. We can calculate the reference discharge following Ponce and Chaganti (1994)

$$Q_R = \frac{Q_b + Q_p}{2} = 2 \text{ m}^3\text{s}^{-1} \tag{16}$$

We then obtain kinematic wave celerity $C$ = 2.157 m s$^{-1}$, and diffusion coefficient $D$ = 50.0 m$^2$ s$^{-1}$. From Eq. (8), with the spatial interval $\Delta x = L$, the time step for the third-order accuracy CPMC is $\Delta t$ = 4637 s, and the Courant number is $C_r = \frac{C \Delta t}{\Delta x} = 1.00$. But with this time step, there are only a few points within range of the simulation time, and much information on temporally and spatially variable discharge would be lost. For easy comparison of the CPMC with the analytical solution, we may choose different time-step sizes, e.g., 1, 2, 5, 10, 20, 50, 100, 200, 500, and 1000 s, and divide the channel into multiple segments ($n_s$). For the second-order accuracy CPMC, we need to keep $C_r$ as close to 1 as possible in our spatial discretization for any specific time-step size (Ponce, 1995, p. 294). For the third-order accuracy CPMC, we divide the channel into multiple segments following Eq. (7). If the channel length is not dividable by the required spatial interval for the third-order accuracy, we would make it as close as practical, and in this case the accuracy would be slightly lower than third order.

From Eq. (15), we have

$$Q_t = \frac{2\pi}{T} \cos\frac{2\pi t}{T} \tag{17}$$

$$Q_x = \frac{2\pi}{L}\cos\frac{2\pi x}{L} \tag{18}$$

and

$$Q_{xx} = -\left(\frac{2\pi}{L}\right)^2 \sin\frac{2\pi x}{L} \tag{19}$$

Incorporating Eqs. (17)–(19) into (1) and simplifying, we obtain the lateral inflow

$$q(x,t) = \frac{2\pi}{CT}\cos\frac{2\pi t}{T} + \frac{2\pi}{L}\cos\frac{2\pi x}{L} + \frac{D}{C}\left(\frac{2\pi}{L}\right)^2 \sin\frac{2\pi x}{L} \tag{20}$$

So our channel routing problem is composed of Eq. (1), initial condition $Q(x,0) = 2 + \sin\frac{2\pi x}{L}$, boundary condition $Q(0,t) = 2 + \sin\frac{2\pi t}{T}$, and lateral inflow calculated by Eq. (20). The channel routing results from second- and third-order accuracy CPMC method are compared with the analytical solution calculated by Eq. (15) at $x=L$. The results of CPMC method with average lateral inflow calculated by Eq. (14) are also compared with the analytical solution. Since the lateral inflow $q(x,t)$ defined by Eq. (20) is not uniformly distributed, we name the method of calculating $\bar{q}_{i+1}^{j+1}$ by Eq. (14) as the simplified method. In the simplified method, we still use the actual values of lateral inflow that are variable with space and time, but the spatial derivatives of lateral inflow are neglected. For uniformly distributed lateral inflow, this simplified method recovers the second-order accuracy.

To calculate $\bar{q}_{i+1}^{j+1}$ in the CPMC using Eq. (11) or (12), we also need the following derivatives of $q(x,t)$

$$q_t = -\frac{1}{C}\left(\frac{2\pi}{T}\right)^2 \sin\frac{2\pi t}{T} \tag{21}$$

$$q_{tt} = -\frac{1}{C}\left(\frac{2\pi}{T}\right)^3 \cos\frac{2\pi t}{T} \tag{22}$$

$$q_x = -\left(\frac{2\pi}{L}\right)^2 \sin\frac{2\pi x}{L} + \frac{D}{C}\left(\frac{2\pi}{L}\right)^3 \cos\frac{2\pi x}{L} \tag{23}$$

$$q_{xx} = -\left(\frac{2\pi}{L}\right)^3 \cos\frac{2\pi x}{L} - \frac{D}{C}\left(\frac{2\pi}{L}\right)^4 \sin\frac{2\pi x}{L} \tag{24}$$

$$q_{3x} = \left(\frac{2\pi}{L}\right)^4 \sin\frac{2\pi x}{L} - \frac{D}{C}\left(\frac{2\pi}{L}\right)^5 \cos\frac{2\pi x}{L} \tag{25}$$

$$q_{4x} = \left(\frac{2\pi}{L}\right)^5 \cos\frac{2\pi x}{L} + \frac{D}{C}\left(\frac{2\pi}{L}\right)^6 \sin\frac{2\pi x}{L} \tag{26}$$

$$q_{xt} = 0 \tag{27}$$

$$q_{xxt} = 0 \tag{28}$$

## 3. Results

The simulated time to peak ($t_p$) by the second- and third-order CPMC methods compare well with the analytical solution of 2500 s for $\Delta t \leq 100$ s (Tables 1 and 2).

| $\Delta t$, s | $n_s$ | $\Delta x$, m | RMSE, m$^3$ s$^{-1}$ | $Q_p$, m$^3$ s$^{-1}$ | $t_p$, s | $C_r$ | $\Delta Q_p$, m$^3$ s$^{-1}$ |
|---|---|---|---|---|---|---|---|
| 1 | 4637 | 2.16 | 9.82E−05 | 2.99995 | 2499 | 1.000 | −4.90E−05 |
| 2 | 2318 | 4.31 | 9.86E−05 | 2.99995 | 2500 | 1.000 | −4.91E−05 |
| 5 | 927 | 10.79 | 1.02E−04 | 2.99995 | 2500 | 1.000 | −4.92E−05 |
| 10 | 464 | 21.55 | 1.12E−04 | 2.99995 | 2500 | 1.001 | −4.97E−05 |
| 20 | 232 | 43.10 | 1.56E−04 | 2.99995 | 2500 | 1.001 | −5.25E−05 |
| 50 | 93 | 107.53 | 4.77E−04 | 2.99992 | 2500 | 1.003 | −8.34E−05 |
| 100 | 46 | 217.39 | 1.66E−03 | 2.99969 | 2500 | 0.992 | −3.07E−04 |
| 200 | 23 | 434.78 | 6.32E−03 | 2.99774 | 2400 | 0.992 | −2.26E−03 |
| 500 | 9 | 1111.11 | 3.99E−02 | 2.97296 | 2000 | 0.970 | −2.70E−02 |
| 1000 | 5 | 2000.00 | 1.39E−01 | 2.98038 | 2000 | 1.078 | −1.96E−02 |

Table 1. Accuracy of the second-order CPMC channel routing with lateral inflow for different time-step sizes.

| $\Delta t$, s | $n_s$ | $\Delta x$, m | RMSE, m$^3$ s$^{-1}$ | $Q_p$, m$^3$ s$^{-1}$ | $t_p$, s | $C_r$ | $\Delta Q_p$, m$^3$ s$^{-1}$ | $\left\| C_r^2 + \dfrac{3}{P_e^2} - 1 \right\|$ |
|---|---|---|---|---|---|---|---|---|
| 1 | 124 | 80.65 | 1.63E−05 | 3.00004 | 2500 | 0.027 | 3.83E−05 | 7.52E−03 |
| 2 | 124 | 80.65 | 1.59E−05 | 3.00004 | 2500 | 0.053 | 3.75E−05 | 5.37E−03 |
| 5 | 123 | 81.30 | 1.52E−05 | 3.00004 | 2500 | 0.133 | 3.59E−05 | 6.57E−03 |
| 10 | 120 | 83.33 | 1.44E−05 | 3.00003 | 2500 | 0.259 | 3.38E−05 | 4.21E−03 |
| 20 | 110 | 90.91 | 1.41E−05 | 3.00003 | 2500 | 0.474 | 3.25E−05 | 5.57E−03 |
| 50 | 74 | 135.14 | 2.65E−05 | 3.00006 | 2500 | 0.798 | 6.32E−05 | 1.01E−02 |
| 100 | 43 | 232.56 | 1.10E−04 | 3.00027 | 2500 | 0.927 | 2.65E−04 | 2.08E−02 |
| 200 | 23 | 434.78 | 6.64E−04 | 2.99956 | 2400 | 0.992 | −4.40E−04 | 1.83E−02 |
| 500 | 9 | 1111.11 | 1.17E−02 | 3.02719 | 2500 | 0.970 | 2.72E−02 | 5.29E−02 |
| 1000 | 5 | 2000.00 | 5.94E−02 | 3.09185 | 2000 | 1.078 | 9.19E−02 | 1.64E−01 |

Table 2. Accuracy of the third-order CPMC channel routing with lateral inflow for different time-step sizes.

For $\Delta t$=200, 500, and 1000 s, the second-order CPMC resulted in smaller $t_p$. The third-order CPMC led to smaller $t_p$ for $\Delta t$=200 and 1000 s. Both methods adequately estimated the peak discharge ($Q_p$, 3 m$^3$ s$^{-1}$).

The RMSE for the third-order CPMC solution is 2–18 times smaller than for the second-order CPMC for each corresponding $\Delta t$ (Table 1 and 2). The RMSE for both methods decreases with $\Delta t$ for $\Delta t \geq 20$ s, and remains nearly constant for $\Delta t < 20$ s (Fig. 2)

For large time steps, the spatial discretizations of second- and third-order accuracy scheme are the same (Tables 1 and 2). For small time steps, however, the third-order accuracy scheme.

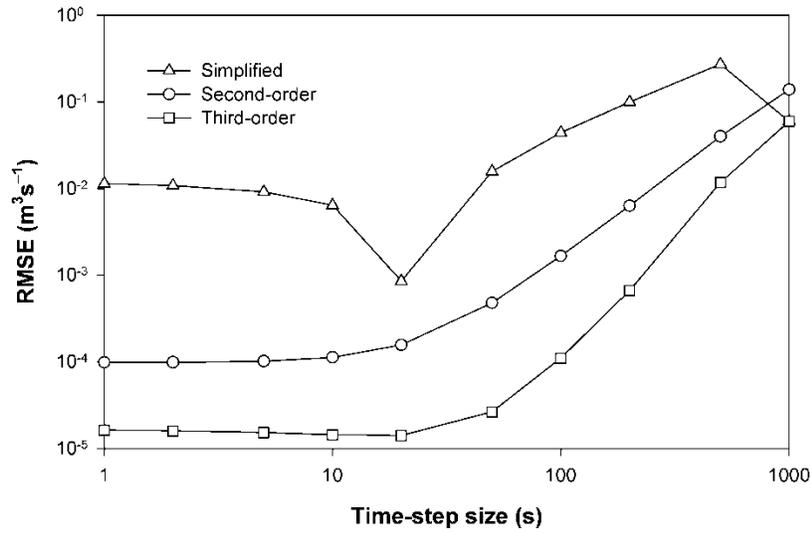

Figure 2. Root-mean-square errors (RMSE) of the simplified, second-, and third-order CPMC.

need fewer spatial steps to reach an improved accuracy than the second-order accuracy scheme, even the third-order accuracy criterion $C_r^2 + \frac{3}{P_e^2} - 1 = 0$ or $C_r = \sqrt{1 - \frac{3}{P_e^2}}$ is not fully met.

For the simplified method, the average lateral inflow are calculated by ignoring the spatial derivatives of lateral inflow but still accounting for different lateral inflow values at different locations. The simulated $Q_p$ was underestimated for $\Delta t \leq 20$ s and overestimated for $\Delta t \geq 50$ s (Table 3). The simulated $t_p$ was comparable with the third-order accuracy scheme for $\Delta t \geq 20$ s but over-estimated for $\Delta t \leq 10$ s.

The RMSE of the simplified method was generally larger than that of the second- or third-order scheme (Fig. 2). The RMSE was smallest when $\Delta t$ was close to 20 s, and increased with decreasing and increasing $\Delta t$. One exception was when $\Delta t=1000$ s, the results were more accurate than when $\Delta t=200$ or 500 s, and were comparable with the third-order accuracy solution. Hence, for this special example, the spatial derivatives of lateral inflow can be neglected if $\Delta t$ and $\Delta x$ were set as 1000 s and 2000 m, respectively.

| $\Delta t$, s | $n_s$ | $\Delta x$, m | RMSE, m³ s⁻¹ | $Q_p$, m³ s⁻¹ | $t_p$, s | $C_r$ | $\Delta Q_p$, m³ s⁻¹ |
|---|---|---|---|---|---|---|---|
| 1 | 4637 | 2.16 | 1.14E−02 | 2.97331 | 2512 | 1.000 | −2.67E−02 |
| 2 | 2318 | 4.31 | 1.09E−02 | 2.97461 | 2512 | 1.000 | −2.54E−02 |
| 5 | 927 | 10.79 | 9.19E−03 | 2.97851 | 2510 | 0.996 | −2.15E−02 |
| 10 | 464 | 21.55 | 6.41E−03 | 2.98499 | 2510 | 1.001 | −1.50E−02 |
| 20 | 232 | 43.10 | 8.51E−04 | 2.99798 | 2500 | 1.001 | −2.02E−03 |
| 50 | 93 | 107.53 | 1.58E−02 | 3.03686 | 2500 | 1.003 | 3.69E−02 |
| 100 | 46 | 217.39 | 4.40E−02 | 3.10319 | 2500 | 0.992 | 1.03E−01 |
| 200 | 23 | 434.78 | 9.97E−02 | 3.23506 | 2400 | 0.992 | 2.35E−01 |
| 500 | 9 | 1111.11 | 2.72E−01 | 3.64472 | 2500 | 0.970 | 6.45E−01 |
| 1000 | 5 | 2000.00 | 5.94E−02 | 3.09185 | 2000 | 1.078 | 9.19E−02 |

Table 3. Accuracy of the CPMC channel routing with simplified calculation of lateral inflow (assuming uniformly distributed) for different time-step sizes.

The largest errors for CPMC solutions of different order of accuracy occur at different times. The largest errors for the second-order CPMC method occur before and after the peak, being overestimates before, and underestimates after, the peak (Fig. 3). The largest error for the third-order or the simplified method occurs only around the peak. The simulation results by the second- and third-order CPMC methods matched the analytical solution well for $\Delta t<500$ s, and by the simplified method for $\Delta t<100$ s.

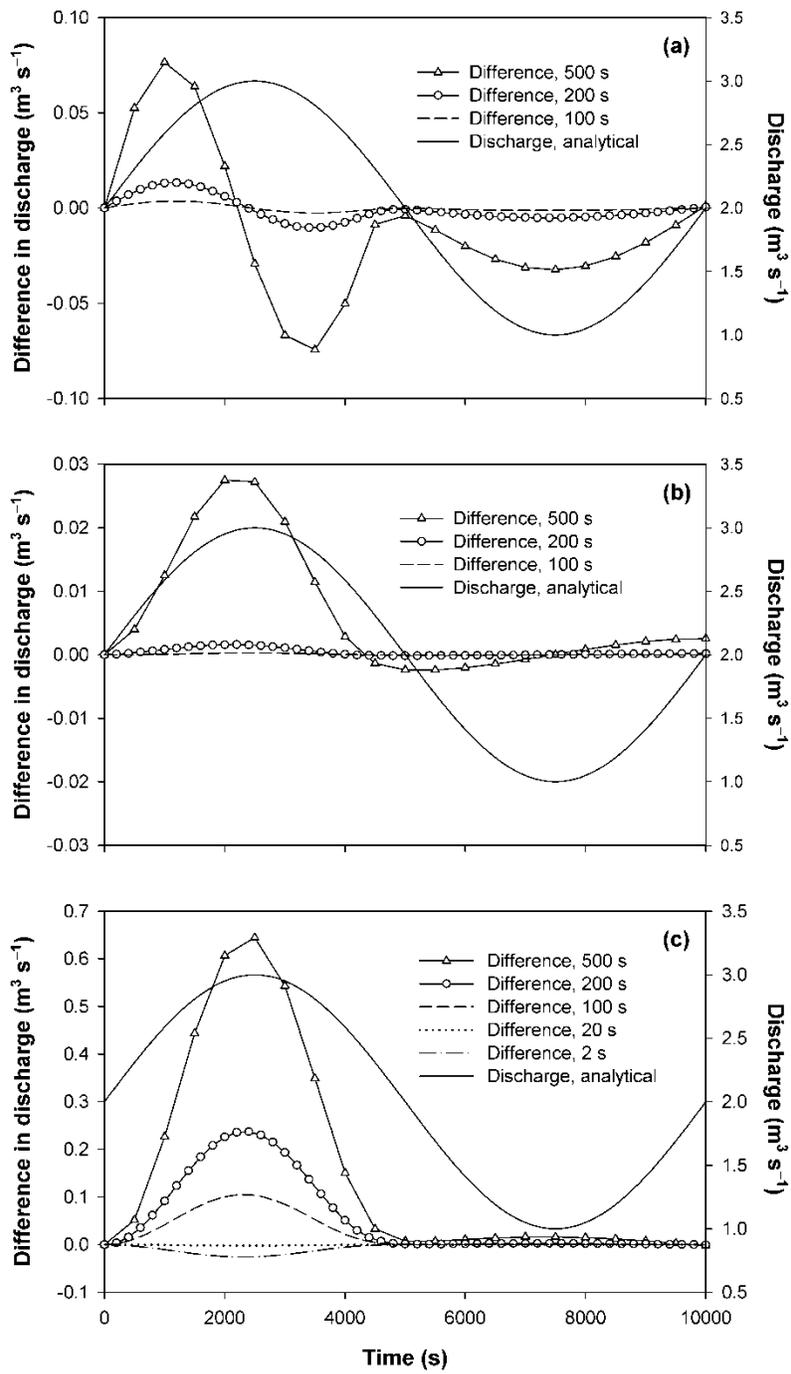

Figure 3. Analytical solution and differences in discharge between numerical and analytical solutions. (a) second-order accuracy CPMC, (b) third-order accuracy CPMC, and (c) simplified method. Note different scales used for difference in discharge in (a), (b), and (c).

## 4. Conclusions

For constant-parameter Muskingum-Cunge diffusion-wave channel routing with spatially and temporally variable lateral inflow, the accuracy of lateral inflow calculation is an important factor affecting the overall channel routing accuracy. In this study, we derived the average lateral inflow term in the second- and third-order accuracy CPMC methods for channel routing. The derived equations indicated that for spatially and temporally variable lateral inflow, the effect of lateral inflow on simulated discharge depended not only on the value of lateral inflow, its spatial and temporal derivatives, the spatial and temporal discretizations, but also on wave celerity and diffusion coefficient of the channel flow.

The second-order CPMC method led to increased accuracy with decreasing time-step sizes, and kept relatively constant for further decreased time-step sizes. Using larger time-step sizes is computationally more efficient, but with higher risk of missing the exact peak discharge point by as much as one time step.

The accuracy of the third-order CPMC solution increased with decreasing time-step sizes, and was higher than the second-order CPMC method, even when the third-order accuracy CPMC method requirement $C_r^2 + \frac{3}{P_e^2} - 1 = 0$ was not fully satisfied because of limitation of constant temporal and spatial intervals used. Its computational costs can be much lower than the second-order CPMC method for smaller time-step sizes when it required few spatial steps. For larger time steps, its spatial discretization became the same as for the second-order scheme. This suggested that for a fixed time step, we can get second-order accuracy CPMC method by maintaining a Courant number of as close to 1 as practical, and obtain a higher accuracy by using a larger spatial step or a smaller Courant number, $C_r = \sqrt{1 - \frac{3}{P_e^2}}$, with the condition that $P_e > \sqrt{3}$.

When we ignore the spatial derivatives of the lateral inflow as in the simplified method, the RMSE of the numerical channel routing results was generally larger than that of the second- and third-order accuracy schemes. It was smallest for a time step of 20 s, and increased with both decreasing and increasing of the time-step size. Only for a special discretization, the simplified method led to the same result with the third-order accuracy scheme.

The second-order accuracy CPMC led to overestimation before and underestimation after, the time of peak discharge. The third-order accuracy CPMC and the simplified method only led to over- or underestimation near the time of peak discharge.

**Acknowledgements**

This work was supported by the USDA CSREES CEAP Grant (No. 2008-48686-04903) and Washington State University.

**Appendix A**

**Derivation of the third-order accuracy CPMC method for constant-parameter diffusion-wave channel routing with lateral inflow**

In order to obtain the $3^{rd}$-order accurate solution of Eq. (2), we calculate the derivatives of $Q$ respect to space and time, and represent them as the derivatives of space only.

First, we rearrange Eq. (1) as

$$Q_t = -CQ_x + DQ_{xx} + Cq \qquad (A1)$$

where,

$$Q_t = \frac{\partial Q}{\partial t}(x,t), \ Q_x = \frac{\partial Q}{\partial x}(x,t), \ Q_{xx} = \frac{\partial^2 Q}{\partial x^2}(x,t), \text{ and } q = q(x,t)$$

are used for brevity, and similar notations are used for the following derivations.

The derivatives are then

$$Q_{tx} = Q_{xt} = -CQ_{xx} + DQ_{3x} + Cq_x \qquad (A2)$$

$$Q_{xxt} = (Q_{xt})_x = -CQ_{3x} + DQ_{4x} + Cq_{xx} \qquad (A3)$$

$$\begin{aligned} Q_{tt} &= -CQ_{xt} + DQ_{xxt} + Cq_t \\ &= -C(-CQ_{xx} + DQ_{3x} + Cq_x) + D(-CQ_{3x} + DQ_{4x} + Cq_{xx}) + Cq_t \\ &= C^2 Q_{xx} - CDQ_{3x} - C^2 q_x - CDQ_{3x} + D^2 Q_{4x} + CDq_{xx} + Cq_t \\ &= C^2 Q_{xx} - 2CDQ_{3x} + D^2 Q_{4x} + Cq_t - C^2 q_x + CDq_{xx} \end{aligned} \qquad (A4)$$

$$Q_{3xt} = (Q_{xxt})_x = -CQ_{4x} + DQ_{5x} + Cq_{3x} \qquad (A5)$$

$$\begin{aligned} Q_{xtt} &= (Q_{xt})_t = -CQ_{xxt} + DQ_{3xt} + Cq_{xt} \\ &= -C(-CQ_{3x} + DQ_{4x} + Cq_{xx}) + D(-CQ_{4x} + DQ_{5x} + Cq_{3x}) + Cq_{xt} \\ &= C^2 Q_{3x} - CDQ_{4x} - C^2 q_{xx} - CDQ_{4x} + D^2 Q_{5x} + CDq_{3x} + Cq_{xt} \\ &= C^2 Q_{3x} - 2CDQ_{4x} + D^2 Q_{5x} + Cq_{xt} - C^2 q_{xx} + CDq_{3x} \end{aligned} \qquad (A6)$$

$$Q_{4xt} = (Q_{3xt})_x = -CQ_{5x} + DQ_{6x} + Cq_{4x} \qquad (A7)$$

$$\begin{aligned}
Q_{3t} &= (Q_{tt})_t = (C^2 Q_{xx} - 2CDQ_{3x} + D^2 Q_{4x} + Cq_t - C^2 q_x + CDq_{xx})_t \\
&= C^2 Q_{xxt} - 2CDQ_{3xt} + D^2 Q_{4xt} + Cq_{tt} - C^2 q_{xt} + CDq_{xxt} \\
&= C^2(-CQ_{3x} + DQ_{4x} + Cq_{xx}) - 2CD(-CQ_{4x} + DQ_{5x} + Cq_{3x}) \\
&\quad + D^2(-CQ_{5x} + DQ_{6x} + Cq_{4x}) + Cq_{tt} - C^2 q_{xt} + CDq_{xxt} \\
&= -C^3 Q_{3x} + C^2 DQ_{4x} + 2C^2 DQ_{4x} - 2CD^2 Q_{5x} - CD^2 Q_{5x} + D^3 Q_{6x} + C^3 q_{xx} \\
&\quad - 2C^2 Dq_{3x} + CD^2 q_{4x} + Cq_{tt} - C^2 q_{xt} + CDq_{xxt} \\
&= -C^3 Q_{3x} + 3C^2 DQ_{4x} - 3CD^2 Q_{5x} + D^3 Q_{6x} + C^3 q_{xx} + Cq_{tt} - C^2 q_{xt} \\
&\quad - 2C^2 Dq_{3x} + CDq_{xxt} + CD^2 q_{4x}
\end{aligned} \tag{A8}$$

Expand the term $Q_{i+1}^{j+1} = Q(x+\Delta x, t+\Delta t)$ in Eq. (2) respect to $(x, t)$ to the third order Taylor series, neglect the superscript $j$ in the derivatives at $(x, t)$ for brevity, and drop the higher order terms,

$$\begin{aligned}
Q_{i+1}^{j+1} &= Q_i^j + Q_x \Delta x + Q_t \Delta t + \frac{1}{2} Q_{xx} \Delta x^2 + \frac{1}{2} Q_{tt} \Delta t^2 + Q_{xt} \Delta x \Delta t \\
&\quad + \frac{1}{6} Q_{3x} \Delta x^3 + \frac{1}{6} Q_{3t} \Delta t^3 + \frac{1}{2} Q_{xxt} \Delta x^2 \Delta t + \frac{1}{2} Q_{xtt} \Delta x \Delta t^2
\end{aligned} \tag{A9}$$

Incorporating the derivative terms and rearranging gives

$$\begin{aligned}
Q_{i+1}^{j+1} &= Q_i^j + Q_x(\Delta x - C\Delta t) + Q_{xx}\left(D\Delta t + \frac{1}{2}\Delta x^2 + \frac{1}{2}C^2 \Delta t^2 - C\Delta x \Delta t\right) \\
&\quad + Q_{3x}\left(-CD\Delta t^2 + D\Delta x \Delta t + \frac{1}{6}\Delta x^3 - \frac{1}{6}C^3 \Delta t^3 - \frac{1}{2}C\Delta x^2 \Delta t + \frac{1}{2}C^2 \Delta x \Delta t^2\right) \\
&\quad + C\Delta t\Bigg[q_i^j + \frac{1}{2}\Delta t(q_t - Cq_x + Dq_{xx}) + \Delta x q_x + \frac{1}{6}\Delta t^2(C^2 q_{xx} + q_{tt} - Cq_{xt} - 2CDq_{3x} \\
&\quad + Dq_{xxt} + D^2 q_{4x}) + \frac{1}{2}\Delta x^2 q_{xx} + \frac{1}{2}\Delta x \Delta t(q_{xt} - Cq_{xx} + Dq_{3x})\Bigg]
\end{aligned} \tag{A10}$$

Similarly,

$$\begin{aligned}
Q_i^{j+1} &= Q_i^j + Q_x(-C\Delta t) + Q_{xx}\left(D\Delta t + \frac{1}{2}C^2 \Delta t^2\right) + Q_{3x}\left(-CD\Delta t^2 - \frac{1}{6}C^3 \Delta t^3\right) \\
&\quad + C\Delta t\left[q_i^j + \frac{1}{2}\Delta t(q_t - Cq_x + Dq_{xx}) + \frac{1}{6}\Delta t^2(C^2 q_{xx} + q_{tt} - Cq_{xt} - 2CDq_{3x} + Dq_{xxt} + D^2 q_{4x})\right]
\end{aligned} \tag{A11}$$

$$Q_{i+1}^j = Q_i^j + Q_x(\Delta x) + Q_{xx}\left(\frac{1}{2}\Delta x^2\right) + Q_{3x}\left(\frac{1}{6}\Delta x^3\right) \tag{A12}$$

Incorporating Eq. (A10), (A11), and (A12) into (2), we have

$$Q_i^j + Q_x(\Delta x - C\Delta t) + Q_{xx}\left(D\Delta t + \frac{1}{2}\Delta x^2 + \frac{1}{2}C^2\Delta t^2 - C\Delta x\Delta t\right)$$

$$+ Q_{3x}\left(-CD\Delta t^2 + D\Delta x\Delta t + \frac{1}{6}\Delta x^3 - \frac{1}{6}C^3\Delta t^3 - \frac{1}{2}C\Delta x^2\Delta t + \frac{1}{2}C^2\Delta x\Delta t^2\right)$$

$$+ C\Delta t\left[q_i^j + \frac{1}{2}\Delta t(q_t - Cq_x + Dq_{xx}) + \Delta x q_x + \frac{1}{6}\Delta t^2(C^2 q_{xx} + q_{tt} - Cq_{xt} - 2CDq_{3x} + Dq_{xxt} + D^2 q_{4x})\right.$$

$$\left. + \frac{1}{2}\Delta x^2 q_{xx} + \frac{1}{2}\Delta x\Delta t(q_{xt} - Cq_{xx} + Dq_{3x})\right]$$

$$= C_1 Q_i^j$$

$$+ C_2\left\{Q_i^j + Q_x(-C\Delta t) + Q_{xx}\left(D\Delta t + \frac{1}{2}C^2\Delta t^2\right) + Q_{3x}\left(-CD\Delta t^2 - \frac{1}{6}C^3\Delta t^3\right)\right.$$

$$\left. + C\Delta t\left[q_i^j + \frac{1}{2}\Delta t(q_t - Cq_x + Dq_{xx}) + \frac{1}{6}\Delta t^2(C^2 q_{xx} + q_{tt} - Cq_{xt} - 2CDq_{3x} + Dq_{xxt} + D^2 q_{4x})\right]\right\}$$

$$+ C_3\left[Q_i^j + Q_x(\Delta x) + Q_{xx}\left(\frac{1}{2}\Delta x^2\right) + Q_{3x}\left(\frac{1}{6}\Delta x^3\right)\right] + C_4 \bar{q}_{i+1}^{j+1}\Delta x$$

(A13)

Equate the coefficient terms related to $Q_i^j$, $Q_x$, $Q_{xx}$, $Q_{3x}$, and $q$, respectively,

$$Q_i^j: \quad 1 = C_1 + C_2 + C_3 \tag{A14}$$

$$Q_x: \quad \Delta x - C\Delta t = C_2(-C\Delta t) + C_3 \Delta x \tag{A15}$$

$$Q_{xx}: \quad D\Delta t + \frac{1}{2}\Delta x^2 + \frac{1}{2}C^2\Delta t^2 - C\Delta x\Delta t = C_2\left(D\Delta t + \frac{1}{2}C^2\Delta t^2\right) + C_3\left(\frac{1}{2}\Delta x^2\right) \tag{A16}$$

$$Q_{3x}: \quad -CD\Delta t^2 + D\Delta x\Delta t + \frac{1}{6}\Delta x^3 - \frac{1}{6}C^3\Delta t^3 - \frac{1}{2}C\Delta x^2\Delta t + \frac{1}{2}C^2\Delta x\Delta t^2$$
$$= C_2\left(-CD\Delta t^2 - \frac{1}{6}C^3\Delta t^3\right) + C_3\left(\frac{1}{6}\Delta x^3\right) \tag{A17}$$

$$q:\quad C\Delta t\left[q_i^j + \frac{1}{2}\Delta t(q_t - Cq_x + Dq_{xx}) + \Delta x q_x + \frac{1}{6}\Delta t^2(C^2 q_{xx} + q_{tt} - Cq_{xt} - 2CDq_{3x} + Dq_{xxt} + D^2 q_{4x})\right.$$
$$\left. + \frac{1}{2}\Delta x^2 q_{xx} + \frac{1}{2}\Delta x \Delta t(q_{xt} - Cq_{xx} + Dq_{3x})\right]$$
$$= C_2 C\Delta t\left[q_i^j + \frac{1}{2}\Delta t(q_t - Cq_x + Dq_{xx}) + \frac{1}{6}\Delta t^2(C^2 q_{xx} + q_{tt} - Cq_{xt} - 2CDq_{3x} + Dq_{xxt} + D^2 q_{4x})\right]$$
$$+ C_4 \bar{q}_{i+1}^{j+1}\Delta x$$

(A18)

Solving the system equations of (A14)–(A16) gives the Muskingum-Cunge coefficients

$$C_1 = \frac{\Delta x + C\Delta t - \frac{2D}{C}}{\Delta x + C\Delta t + \frac{2D}{C}} \tag{A19}$$

$$C_2 = \frac{-\Delta x + C\Delta t + \frac{2D}{C}}{\Delta x + C\Delta t + \frac{2D}{C}} \tag{A20}$$

and

$$C_3 = \frac{\Delta x - C\Delta t + \frac{2D}{C}}{\Delta x + C\Delta t + \frac{2D}{C}} \tag{A21}$$

The coefficients are the same as that given by Chow et al. (1988) and Ponce (1995). And it shows that the CPMC method without further restriction is second-order accurate.

Incorporating Eq. (A20) and (A21) into (A17) and simplifying, we have

$$C^4 \Delta t^2 - C^2 \Delta x^2 + 12D^2 = 0 \tag{A22}$$

solving for $\Delta x$, we have

$$\Delta x = \sqrt{C^2 \Delta t^2 + \frac{12D^2}{C^2}} \tag{A23}$$

or solving for $\Delta t$, we have

$$\Delta t = \frac{1}{C}\sqrt{\Delta x^2 - \frac{12D^2}{C^2}} \qquad (A24)$$

Eqs. (A23) and (A24) are the relationships between $\Delta x$ and $\Delta t$ required to maintain the third-order accurate for the CPMC method, and it has been derived by Bajracharya and Barry (1997) and Szel and Gaspar (2000) for CPMC method solving the diffusion-wave channel routing without lateral inflow.

Dividing both sides of (A22) by $C^2\Delta x^2$ and introducing $C_r$ and $P_e$, we can simplify (A22) to a dimensionless equation required for eliminating the dispersion error to obtain the third-order CPMC method (Szel and Gaspar, 2000): $C_r^2 + \frac{3}{P_e^2} - 1 = 0$.

From Eq. (A24), in order for $\Delta t$ to be real, we must have

$$\Delta x > \frac{2\sqrt{3}D}{C} \qquad (A25)$$

From Eq. (A18), we obtain

$$C_4 \bar{q}_{i+1}^{j+1} \Delta x = \frac{2C\Delta t \Delta x}{\Delta x + C\Delta t + \frac{2D}{C}}\left[ q^j + \frac{1}{2}q_t\Delta t + \frac{1}{2}q_x\left(\Delta x + \frac{2D}{C}\right) + \frac{1}{6}q_{tt}\Delta t^2 + \frac{1}{4}q_{xx}\left(\Delta x^2 - \frac{1}{3}C^2\Delta t^2 + \frac{2D\Delta x}{C}\right) \right.$$
$$\left. + \frac{1}{4}q_{xt}\left(\Delta x + \frac{1}{3}C\Delta t + \frac{2D}{C}\right)\Delta t + \frac{1}{4}q_{3x}\left(\Delta x - \frac{1}{3}C\Delta t + \frac{2D}{C}\right)D\Delta t + \frac{1}{6}q_{xxt}D\Delta t^2 + \frac{1}{6}q_{4x}D^2\Delta t^2 \right]$$

$$(A26)$$

Letting

$$C_4 = \frac{2C\Delta t}{\Delta x + C\Delta t + \frac{2D}{C}} \qquad (A27)$$

we get

$$\bar{q}_{i+1}^{j+1} = q^j + \frac{1}{2}q_t\Delta t + \frac{1}{2}q_x\left(\Delta x + \frac{2D}{C}\right) + \frac{1}{6}q_{tt}\Delta t^2 + \frac{1}{4}q_{xx}\left(\Delta x^2 - \frac{1}{3}C^2\Delta t^2 + \frac{2D\Delta x}{C}\right)$$
$$+ \frac{1}{4}q_{xt}\left(\Delta x + \frac{1}{3}C\Delta t + \frac{2D}{C}\right)\Delta t + \frac{1}{4}q_{3x}\left(\Delta x - \frac{1}{3}C\Delta t + \frac{2D}{C}\right)D\Delta t + \frac{1}{6}q_{xxt}D\Delta t^2 + \frac{1}{6}q_{4x}D^2\Delta t^2$$

(A28)

And incorporating Eq. (A20) into (A27) and simplifying results in

$$C_4 = \frac{2C\Delta t}{\Delta x + C\Delta t + \frac{2D}{C}} = C_1 + C_2 = 1 - C_3 \tag{A29}$$

Eqs. (2), (A19)–(A21), (A29), and (A28) together form the third-order CPMC method with spatial or temporal limitations defined by Eq. (A23) and (A24), respectively.

If the spatial variation of lateral inflow is negligible so that the derivatives of $q$ with respect to $x$ vanish in Eq. (A28), the average lateral inflow can also be estimated simply from a discrete dataset. Since (Thomas, 1995)

$$q_t = \begin{cases} \dfrac{-q_i^{j+2} + 4q_i^{j+1} - 3q_i^j}{2\Delta t} + O(\Delta t^2) & \text{for } j = 0 \\ \dfrac{q_i^{j+1} - q_i^{j-1}}{2\Delta t} + O(\Delta t^2) & \text{for } j = 1,2,\text{K} \end{cases} \tag{A30}$$

and

$$q_{tt} = \begin{cases} \dfrac{q_i^{j+2} - 2q_i^{j+1} + q_i^j}{\Delta t^2} + O(\Delta t) & \text{for } j = 0 \\ \dfrac{q_i^{j+1} - 2q_i^j + q_i^{j-1}}{\Delta t^2} + O(\Delta t^2) & \text{for } j = 1,2,\text{K} \end{cases} \tag{A31}$$

Incorporating Eqs. (A30) and (A31) into (A28) and simplifying, we have

$$\bar{q}_{i+1}^{j+1} = \begin{cases} \dfrac{-q_i^{j+2} + 8q_i^{j+1} + 5q_i^j}{12} + O(\Delta t^3) & \text{for } j = 0 \\ \dfrac{5q_i^{j+1} + 8q_i^j - q_i^{j-1}}{12} + O(\Delta t^3) & \text{for } j = 1,2,\text{K} \end{cases} \tag{A32}$$

Following the same procedure, the second-order accuracy lateral inflow term can be obtained as

$$\bar{q}_{i+1}^{j+1} = q^j + \frac{1}{2}q_t\Delta t + \frac{1}{2}q_x\left(\Delta x + \frac{2D}{C}\right) + \frac{1}{2}q_{xx}D\Delta t \qquad (A33)$$

If we ignore the spatial derivatives, and use $q_t = \frac{q_i^{j+1} - q_i^j}{\Delta t} + O(\Delta t)$, Eq. (A33) can be simplified to (Chow et al., 1988)

$$\bar{q}_{i+1}^{j+1} = \frac{q_i^{j+1} + q_i^j}{2} \approx \frac{q_{i+1}^{j+1} + q_{i+1}^j}{2} + O(\Delta t^2) \quad \text{for } j = 0,1,2\text{K} \qquad (A34)$$

For kinematic wave channel routing, $D = 0$, Eq. (A33) is simplified to

$$\bar{q}_{i+1}^{j+1} = q^j + \frac{1}{2}q_t\Delta t + \frac{1}{2}q_x\Delta x \qquad (A35)$$

If we estimate the derivatives $q_x$ and $q_t$, respectively, by $q_x = \frac{q_{i+1}^j - q_i^j}{\Delta x} + O(\Delta x)$ and $q_t = \frac{q_i^{j+1} - q_i^j}{\Delta t} + O(\Delta t)$, Eq. (A35) becomes

$$\bar{q}_{i+1}^{j+1} = \frac{q_{i+1}^j + q_i^{j+1}}{2}. \qquad (A36)$$